\documentstyle[aps,pre,epsf]{revtex}
\begin{document}

%\advance\textheight by 0.2in
\draft
%\twocolumn[\hsize\textwidth\columnwidth\hsize\csname@twocolumnfalse%
%\endcsname

\title{Scaling Laws and Topological Exponents in Voronoi Tessellations\\
       of Intermittent Point Distributions }
\author{Haye Hinrichsen and Gudrun Schliecker}
\address{Max-Planck-Institut f\"ur Physik Komplexer Systeme\\
         N\"othnitzer Stra{\ss}e 38, 01187 Dresden, Germany}
\date{J. Phys. {\bf A 31}, L451 (1998)}
\maketitle

\begin{abstract}
Voronoi tessellations of scale-invariant fractal sets are
characterized by topological and metrical properties that are 
significantly different from those of natural cellular structures.
As an example we analyze Voronoi diagrams of intermittent
particle distributions generated by a directed
percolation process in $2+1$ dimensions.
We observe that the average area of a cell 
increases much faster with the number of its neighbours
than in natural cellular structures where
Lewis' law predicts a linear behaviour.
We propose and numerically verify
a universal scaling law that relates shape and size of the
cells in scale-invariant tessellations. An
exponent, related to the topological properties of the
tessellation, is introduced and estimated numerically.
\end{abstract}

\pacs{{\bf PACS numbers:} 68.90.+g, 64.60.Ak, 82.70.Rr \\
      {\bf Key words:} \hspace{6mm} Mosaics, Voronoi construction,
      directed percolation}
%] %this closed bracket is for twocolumn style
%
% Explanation of PACS numbers:
% 68.90.+g:     Topics in structure, surfaces, interfaces and thin films
% 64.60.Ak:     Fractal and percolation studies
% 82.70.Rr:     Aerosols and foams
%
%

\vspace{2mm}

Planar random mosaics are often encountered in nature, as, for example,
in cuts of biological tissues~\cite{tissuesold,tissuesnew} and 
two-dimensional soap froths~\cite{soap,flyvbjerg}.
Thorough investigations of the topological and metrical
properties of a huge variety of natural mosaics show a surprising similarity
despite of the fact that the molding forces are completely 
different~\cite{similar}. 

The Voronoi construction~\cite{voronoi}
allows to generate a mosaic from a set of 
arbitrarily distributed seeds. It assigns to each seed a cell which is
defined as the set of all points of the plane which are at least as close
to this seed as to any other seed.  Since the resulting Voronoi tessellation 
is a space-filling cellular structure it allows to introduce the
notion of neighbourhood, i.e., two seeds are neighboured if their cells
share at least one side. Thus the Voronoi construction enables us to
investigate the neighbour statistics by analyzing the topological
properties of the corresponding tessellation.
This approach has been successfully applied to
experimental monosize disc assemblies on an air 
table~\cite{airtable1,airtable2},
confirming the universality of random mosaics. In addition, 
the Voronoi construction offers 
an alternative tool for the characterization 
of the order-disorder transition
which occurs when the density is reduced. 

So far all experimentally investigated cellular structures are
characterized by a typical scale, namely the mean size of a cell.
The question therefore arises how the mentioned universal
laws are affected in the case of {\it scale-invariant} mosaics.
In scale-invariant tessellations one expects that the area 
distribution of the cells exhibits a power-law behaviour.
Moreover, it was shown in Ref.~\cite{fractalfoam} that in
Sierpinski cellular structures the distribution of
edge numbers also behaves algebraically.
In the present work we are particularly interested in 
the relation between topological and metrical properties of
scale-invariant mosaics. For this purpose we investigate 
Voronoi tessellations of intermittent 
particle distributions with algebraic long-range
correlations that are generated by planar stochastic point processes.

As an example we consider a Directed Percolation (DP) process
in $2+1$ dimensions~\cite{percol,grass89}. In DP --
interpreted as a time-dependent stochastic process --
particles on a lattice either produce an offspring or self-destruct.
Depending on the rates of offspring production and self-destruction
this process exhibits a phase transition from a fluctuating active
phase into an absorbing state without particles from where the 
system cannot escape. We use directed bond percolation~\cite{percol}
which is controlled by a single parameter, namely the percolation
probability $p$. Below a critical threshold $p<p_c$ 
the system approaches the absorbing 
state in exponentially short time whereas in the 
active phase with $p>p_c$ a fluctuating
stationary state exists on the infinite lattice. Close to criticality
such a stationary DP process evolves through configurations that
are characterized by spatially intermittent patterns with 
long-range correlations~\cite{henkel},
i.e., very few particles form highly localized clouds separated by
large empty regions.

\begin{figure}
\epsfxsize=145mm     
\centerline{ \epsffile{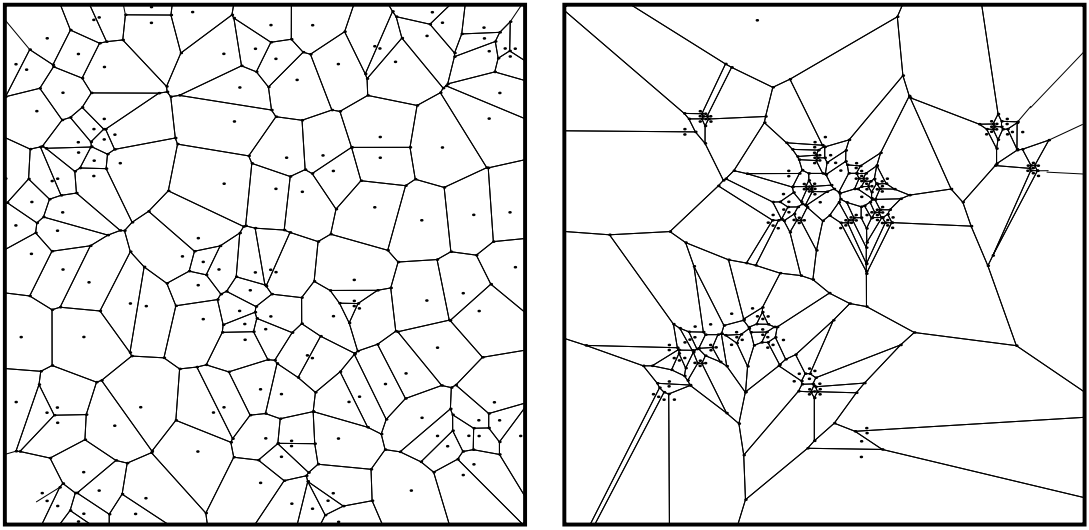}  } 
\vspace{4mm}
\caption {
Ordinary and scale-invariant cellular structures. The figure
shows parts of Voronoi tessellations for a Poissonian distribution of
points (left) and for a point set generated by an almost critical
DP process (right).
}
\label{FigureVergleich}
\end{figure}

In the present work we use spatial configurations (snapshots) 
of active particles generated by an 
almost critical stationary DP process as point sets for a Voronoi 
construction and to study the properties of the resulting tessellations.
As shown in Fig.~\ref{FigureVergleich}, such a tessellation appears to
be very different from natural cellular structures. The reason is that
point sets generated by an almost critical DP process 
approximate a fractal set, apart from lower 
and upper cutoffs due to system size and lattice spacing. 
In contrast to natural cellular structures the 
corresponding Voronoi tessellations are invariant under rescaling.
Here we address the question how the topological and metrical properties of 
scale-invariant tessellations differ from those of natural
structures. To this end we focus on the correlation between
shape and size of the cells.

The simplest quantity describing these correlations is
$\langle A \rangle_k / \langle A \rangle$, the normalized average 
area of an arbitrarily chosen $k$-sided cell. As pointed out first
by Lewis~\cite{tissuesold}, $\langle A \rangle_k$ increases {\it linearly}
with $k$ for a huge variety of tessellations. Deviations from Lewis'
law have so far been observed in soap froth 
experiments~\cite{soap,flyvbjerg} and
air table tessellations in the dense packing regime~\cite{airtable2}.
These deviations, however, are restricted to cells with only
few neighbours whereas for large $k$ the linear law appears
to be asymptotically valid~\cite{flyvbjerg}.

%
% -----------------------------Figure Vergleich------------------------------
%

In the present work the distribution $P(k,A)$ of $k$-sided
cells with area $A$ generated by an almost critical
DP process is studied numerically. To this end a 
directed bond percolation process with parallel updates is
simulated on a $500 \times 500$ square lattice with periodic
boundary conditions. The simulations are performed in the
active phase with a small reduced percolation probability
$p-p_c = 10^{-4}$, using the critical percolation threshold
$p_c=0.287338$~\cite{pc}. 
After equilibration over $10^{5}$ time steps 
the system evolves in a practically
stationary state with on average about $10^3$ particles. 
The value of $p-p_c$ and the large lattice
size ensure that the systems stays in the active phase.
At intervals of $100$ time steps
the actual configuration of active particles is converted
into the corresponding Voronoi tessellation~\cite{LEDA}.
Averaging over $1000$ of such tessellations we 
determine the relative frequency $P(k,A)$  
as well as the mean area $\langle A \rangle_k$ of a $k$-sided cell.

%
% ------------------------------Figure Lewis -------------------------------
%
\begin{figure}
\epsfxsize=180mm     
\centerline{ \epsffile{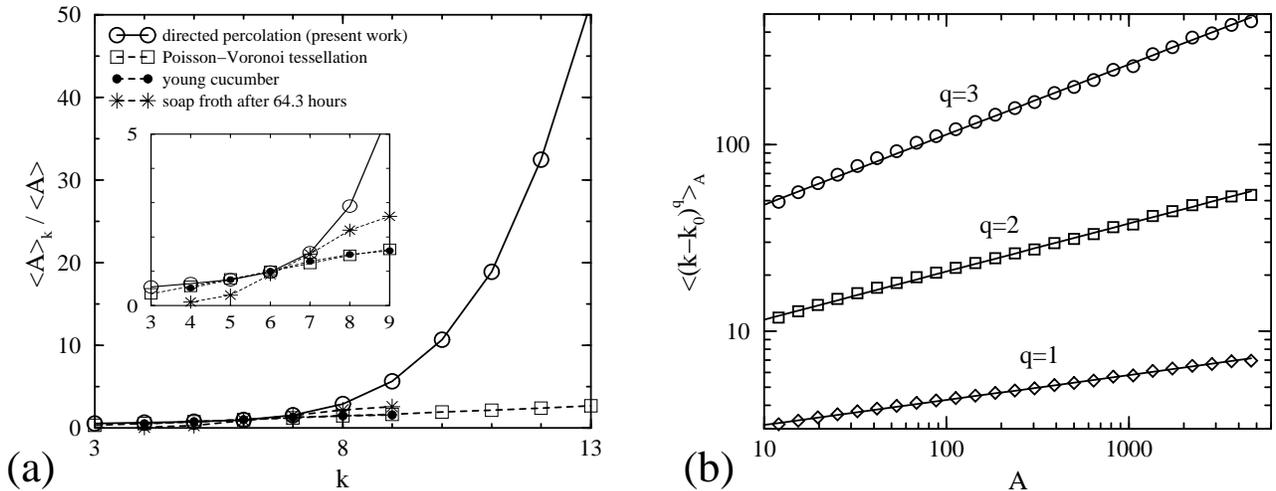}  } 
\caption {
(a) Normalized average area $\langle A \rangle_k/
\langle A \rangle$ of $k$-sided cells
generated by a DP process in comparison with 
a Poisson distribution~[15], 
a cucumber~[1], and a soap froth~[3].
The inset shows a magnification for small values. 
(b) First three moments $\langle (k-k_0)^q \rangle_A$ of
the edge number distribution for cells with area $A$.
The shift $k_0=3$ is explained in the text.
}
\label{FigureLewis}
\end{figure}

%
% -------------------------------Figure Results------------------------------
%
\begin{figure}
\epsfxsize=175mm   
\centerline{ \epsffile{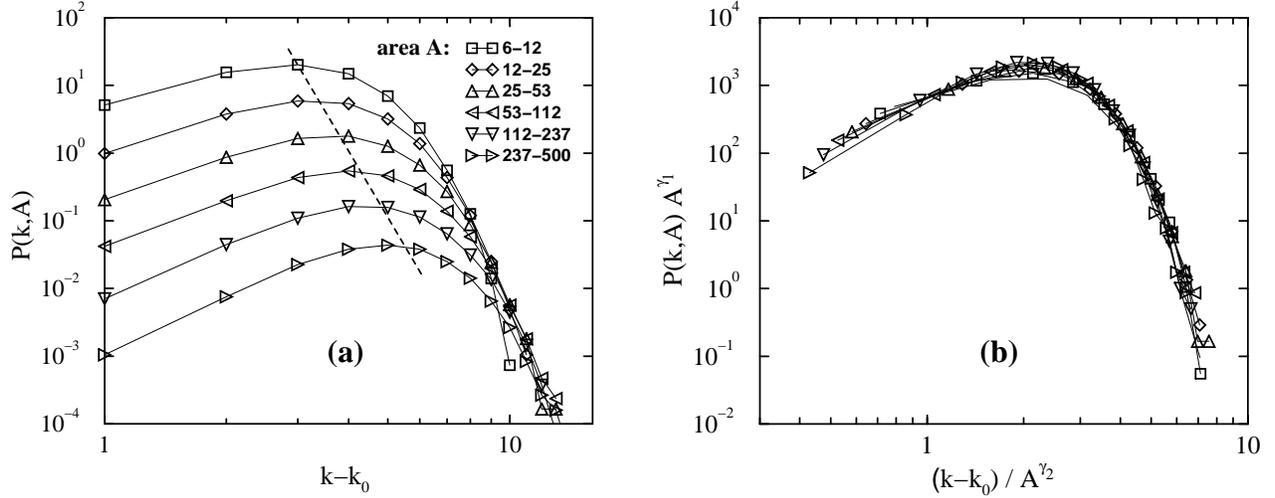}  } 
\caption {
Numerical results from Monte-Carlo simulations in
the scaling regime. (a) Double-logarithmic
representation of the relative frequency $P(k,A)$ of cells with
$k$ edges and area $A$ in units of squared lattice spacing.
The dashed line indicates the location of the maxima. 
(b) Collapse of the same data using the
scaling form~(\ref{ScalingForm})visualizing
the scaling function $\Phi$. The best collapse is 
obtained for the topological exponent $\tau=0.20(4)$.
}
\label{FigureResults}
\end{figure}

As shown in Fig.~\ref{FigureLewis}a, the curve for 
$\langle A \rangle_k / \langle A \rangle$ shows an
untypical behaviour as it increases significantly faster than in 
cellular structures with a typical scale~\cite{poisson,tissuesold,soap}.
It neither resembles Lewis' linear law nor a quadratic behaviour
that has been discussed in the context of a perimeter law
for metallurgical grain structures~\cite{flyvbjerg}.
The numerical results instead suggest an algebraic increase.
In order to check the quality of the scaling, we measured the
first three moments of the edge number distribution of cells
with area $A$ (see Fig.~\ref{FigureLewis}b). We observed that
the best results are obtained if the edge numbers are measured
with respect to their natural minimum $k_0=3$. The slopes are
estimated by $0.13(2)$, $0.26(3)$, and $0.38(5)$ for
$q=1,2,3$, respectively, indicating simple scaling.

Our measurements for $P(k,A)$ are shown 
in Fig.~\ref{FigureResults}a. As $A$ increases, the maximum of
the curves is shifted along the dashed line 
to higher edge numbers $k$. The 
observation that -- despite of the discreteness of $k$ -- all curves
roughly have the same shape supports the conjecture that
the cell topology in DP may exhibit universal scaling.

Let us recall the scaling properties of DP. 
As usual, we denote the directed dimension (time) 
by the index ${||}$ and the other spatial dimensions by $\perp$.
In the active phase close to criticality
the stationary particle intensity $\rho$ 
and the spatial and temporal correlation lengths 
$\xi_\perp$, $\xi_{||}$ scale as~\cite{percol}
\begin{equation}
\label{UsualScaling}
\rho \sim (p-p_c)^\beta\,,
\quad
\xi_\perp \sim (p-p_c)^{-\nu_\perp}\,,
\quad
\xi_{||} \sim (p-p_c)^{-\nu_{||}}\,.
\end{equation}
The exponents $\beta$, $\nu_\perp$, and $\nu_{||}$ are the three
basic critical exponents of DP which usually determine other 
DP exponents by simple scaling relations. In $2+1$ dimensions they
have been numerically estimated~\cite{grass89} by
$\beta=0.584$, $\nu_\perp=0.734$, and $\nu_{||}=1.295$.

The Voronoi construction maps each active site onto a cell with a
certain area $A$ and a topology $k$. The cell topology, i.e.,
the number of neighbours, can be regarded as an additional quantity for the
description of a DP process in more than one spatial dimension.
In order to find an appropriate scaling relation, we conjecture -- in
analogy to Eq.~(\ref{UsualScaling}) -- that the cell topology
exhibits scaling properties similar to those of
distances and densities. This means that the 
corresponding scaling regime 
is restricted to $k < \kappa$, where $\kappa$
diverges close to the transition as
\begin{equation}
\label{TopologyScaling}
\kappa \sim (p-p_c)^{-\tau}\,.
\end{equation}
Here $\kappa$ limits the scaling regime of edge numbers similarly
as $\xi_\perp$ limits the scaling regime of distances in the
active phase. $\tau$ is a critical exponent which will be
referred to as {\it topological exponent} of DP.
At present it is not clear whether $\tau$ is a novel exponent or
related to the bulk exponents $\beta, \nu_\perp, \nu_{||}$ by
a scaling relation (as, for example, the critical initial slip
exponent of DP~\cite{crslip}). 

In the scaling regime we expect the probability 
distribution $P(k,A)$ for cells with topology $k$
and area $A$ to obey the scaling form
\begin{equation}
\label{RawScalingForm}
P(k,A) \sim A^{-\gamma_1}\,\Phi(k A^{-\gamma_2}) 
\,,
\qquad \qquad
(k < \kappa, A<\xi_\perp^2)
\end{equation}
where $\Phi(z)$ is a universal scaling function. The exponents
$\gamma_1,\gamma_2$ can be derived as follows:
Summing over the edge numbers $k$ we obtain the area distribution
\begin{equation}
P(A) = \sum_k P(k,A) \simeq
\int_0^\infty dk\,P(k,A) \sim A^{\gamma_2-\gamma_1} \,.
\end{equation}
Assuming that $0<\gamma_2-\gamma_1+2 < 1$ we can compute
the average area of cells for a given scaling length $\xi_\perp$
by
\begin{equation}
\langle A \rangle \sim \int_{a^d}^{\xi_\perp^d} dA\,A\,P(A) \sim
        \xi_\perp^{d(\gamma_2-\gamma_1+2)}  \,,
\end{equation}
where $a$ is the lattice spacing and $d=2$ the spatial dimension.
Since $\langle A \rangle$ is inversely proportional to the
intensity of points $\rho \sim \xi_\perp^{-\beta/\nu_\perp}$
we obtain the relation $\gamma_1=\gamma_2+2-\beta/d\nu_\perp$
by comparing the exponents. On the
other hand, the argument of the scaling 
function $\Phi$ in Eq.~(\ref{RawScalingForm}) should be invariant 
under rescaling 
$A \rightarrow b^d A, k \rightarrow b^{\tau/\nu_\perp} k$, 
leading to $\gamma_2=\tau/d\nu_\perp$. Thus the expected scaling
relation reads
\begin{equation}
\label{ScalingForm}
P(k,A) \sim A^{2+(\tau-\beta)/d\nu_\perp}\,\Phi(k A^{-\tau/d\nu_\perp}) 
\,,
\qquad \qquad
(k < \kappa, A<\xi_\perp^2)
\end{equation}
In order to verify this scaling law, 
we numerically estimated the topological exponent~$\tau$ 
by data collapse (see Fig.~\ref{FigureResults}b).
Again, the best collapse is obtained
if one replaces $k$ by $k-k_0$
with $k_0=3$, which is the natural minimum 
for the number of sides of a cell.
Our best estimate is $\tau=0.20(4)$, corresponding
to $\gamma_1=1.73(3)$ and $\gamma_2=0.13(3)$.
The collapse is fairly convincing, supporting the above
scaling hypothesis. In addition, the result is in
agreement with the measurements in Fig.~\ref{FigureLewis}b.
We note that this result should not depend 
on the specific choice of the DP dynamics (e.g., bond or site
percolation). Rather we expect $\tau$ to be a
universal exponent characterizing the DP 
class in $2+1$ dimensions. The question whether $\tau$ is independent
or related to the other scaling exponents is still open.

It should be noted that the proposed scaling hypothesis needs further
verification as the scaling regime of $k$ extends over
one decade only. The main limitation is the small 
value of $\tau$, i.e., each decade of scaling range in $k$
requires about five decades of scaling range in $A$.
However, it can be shown that Eq.~(\ref{RawScalingForm})
holds exactly in the case of a simple Sierpinski gasket
which strongly supports our scaling hypothesis~\cite{gudrun}.
The scaling form assumes that the underlying
point distribution is a simple fractal. 
This is the case for spatial cuts of critical DP states,
as stated in Ref.~\cite{henkel}. 
For multifractal cellular structures, however, the proposed 
scaling relation should be replaced
by appropriate multiscaling laws~\cite{gudrun}.

We thank W. Wolf for teaching us how to use the
LEDA software package.


\begin{thebibliography}{99}
\bibitem{LEDA}
                The Voronoi construction was performed with help
                of the LEDA software package. We extended the
                algorithm in a way that periodic boundary
                conditions are taken in to account.
\end{thebibliography}
\end{document}